# Effect of Informational Interventions on EV Adoption Intention: Evidence from a Tier II City in India

Authors: Pranshu Raghuvanshi[1] and Anjula Gurtoo[1]

## Abstract

This study investigates the effectiveness of targeted informational interventions on electric vehicle adoption intention. A randomised controlled field experiment with three treatment groups and a control group was used to study the effectiveness of three informational interventions. Participants in each treatment group received a distinct informational intervention: cost-based, range-based, and norm-based. Two of the three interventions (range-based and norm-based), designed to reduce behavioural and psychological barriers, were found to be significant. The cost-based intervention was not significant, suggesting that financial motives alone may not be sufficient to lead to an increase in the adoption of electric vehicles. The significant effect observed for the range-based and norm-based interventions suggests that the discomfort related to the technology must be addressed, and social norms can be effectively utilised to promote electric vehicles at low cost. Although adoption is not guaranteed with self-reported intentions, the findings suggest that carefully framed informational interventions guide behavioural intentions towards sustainable technologies. The most significant contribution of the study is to the literature on demand-side policy instruments, which suggests that financial incentives can be complemented by other informational interventions to accelerate the adoption of sustainable mobility.

[1] Indian Institute of Science, Bangalore, India

**Introduction**

The targets set in the Paris Agreement, to limit the increase in Earth's average temperature to 2 degrees Celsius and preferably to 1.5 degrees Celsius compared to pre-industrial levels by the end of this century, are on the verge of being breached. The World Meteorological Organisation (WMO), in Global Gas Bulletin 2023, reported that $CO_2$ emissions in the atmosphere are 50 per cent higher than pre-industrial levels (WMO, 2023). WMO also reported an 80 per cent probability that the 1.5 degrees Celsius threshold will be temporarily breached between 2024 and 2028 (WMO, 2024). Similar observations were also made by a report published by IPCC in 2023 (Lee, 2023). The Copernicus observatory in Europe confirmed these apprehensions. It was found that a 1.5 degrees Celsius threshold was breached in 2024, and it was the hottest year recorded on Earth (Mooney, Tauschinski, & Bernard, 2025). The Paris Agreement also mentioned that to contain the rise in the Earth's average temperature within 1.5 degrees Celsius, emissions must peak by 2025 and decline substantially by 2030 (UN). However, we are nowhere close to fulfilling this requirement. With thresholds already temporarily breached, there is a need to take measures to control emissions quickly.

The transport sector is responsible for 23 per cent of the $CO_2$ emissions globally (IEA, 2023). Most of it comes from the road transport sector. According to the IEA, the road sector accounts for 90 per cent of transport emissions. The emissions from the road sector, also known as the Traffic-Related Air Pollution (TRAP) (Palinkas, et al., 2025), can be reduced by using alternative fuels like electric vehicles (EVs). The introduction and adoption of EVs in Europe have decreased TRAP significantly (EEA, 2018), especially in cities like Bergen (once the worst air quality). Norway has improved its quality and become one of the best in Europe (Hanley, 2018). EVs' share in total sales was 20 per cent in Europe in 2024 (IEA, 2025). However, when we look at the developing countries, the progress is not that encouraging. In 2024, the EV penetration in the private car segment in India stood at 7.46 per cent (Kumar, 2025). The Indian government set a target to achieve EV sales penetration of 30 per cent by 2030 under the FAME (Faster Adoption of Electric Vehicles) scheme. The scheme was launched in 2015. Given the time frame of 10 years, the progress on this front is slower than anticipated.

While most studies have focused on the effect of EVs' instrumental attributes (price, range, charging infrastructure, maintenance shops, etc.) (Patel, Singh, Arora, & Mahapatra, 2024; Patel, et al., 2024; Singh & Paul, 2025), only a few have related their slow adoption to the consumer perception of EVs. We see this slow adoption as arising from information asymmetry and the inertial thinking of the consumers (Zhang, Shanyong, Wan, Zhang, & Zhao,

2022; Stampatori & Rossetto, 2024). Consumer perceptions have not changed much, once shaped by the initial information they obtained about EVs. Hence, it is essential to provide them with accurate information about advancements in the instrumental attributes of EVs. This information should be simplified enough for them to make sense.

In this study, we test three informational interventions on consumers' perceptions of EVs. Two of the three interventions provide information on the EV's instrumental attributes, and the third intervention uses social norms to change consumers' perceptions of EVs. Intervention 1 focuses on the monetary benefits of electric vehicles. It gives information on the low operational cost of EVs (fuel savings and low maintenance costs) and the economic value of incentives offered by the government while purchasing the EV. The target is to inform consumers that the low operational cost can offset the high upfront cost of buying an EV. We provide information on the expected time to break even on their EV purchase. While Intervention 1 focuses on the costing aspect of the electric vehicle, the focus of Intervention 2 is to reduce range anxiety. We provide information to consumers on improvements in the EV range, the status of the charging infrastructure, and the fast-charging features of electric vehicles.

The third intervention uses social norms to influence the perception of the consumers. The consumers are provided with information on the story of a businessman from their region, who is highly respected in his community for his sustainable choices and is seen as someone who leads the change. The story also mentions the satisfaction (financial and social) he derives from adopting the electric vehicle. Using social norms for behaviour change has been extensively explored in behavioural public policy (Demarque, Charalambides, Hilton, & Waroquier, 2015; Farrow, Grolleau, & Ibanez, 2017). Farrow et al. provide a detailed account of studies on social norms intervention on pro-environmental behaviour. Evidence-based studies demonstrate the effect of social norms on a wide range of pro-environmental behaviour, including conservation activities (Ferraro, Miranda, & Price, 2011), energy efficiency (Allcott, 2011), recycling (Schultz, 1999), reduce and reuse (Bohner & Schlueter, 2014), etc.

This paper is organised as follows. Section 2 discusses the relevant literature on randomised controlled trials and informational interventions made in this study. The methodology employed in this study is discussed in Section 3, followed by the results in Section 4. These results and their implications are discussed in Section 5. Section 6 contains the conclusion and outlines the scope for future research.

**Literature Review**

The assumption of rationality in classical economics is challenged when we consider behavioural aspects (Kahneman, 2003). Individuals face several constraints in decision-making. These constraints are both cognitive and non-cognitive. While cognitive constraints refer to the limited ability of the human mind to store and process information (Kahneman, 2011), non-cognitive constraints are external to the cognitive skills of the decision maker, such as information asymmetry and peer influence (Thaler & Sunstein, 2008). Cognitive constraints may refer to individuals' inability to understand and use the available information. The unavailability of the critical information in the correct quantity at the right time can be understood as a non-cognitive constraint. These constraints shape an individual's perception, reason and ability to make decisions, leading to biases and errors in judgment (Tversky & Kahneman, 1974). This study contributes to the literature on economics, decision-making, and sustainability by examining the effect of informational interventions that aim to either correct cognitive impediments or exploit them to promote sustainable behaviour.

While studying individuals' decision-making, Herbert Simon argued that individuals use values (impulses, guesswork, heuristics, peer advice, etc.) rather than facts when faced with constraints in decision-making. These constraints can be cognitive or non-cognitive due to information shortage and overload (Simon, 1947). Potential buyers of the electric vehicles may face information asymmetry and cognitive constraints. They may not be able to evaluate the EV's lifetime savings because of the limited information (Allcott, 2013) that they possess about EVs (running cost, maintenance cost, difference when compared to a conventional vehicle) or their limited ability to process that information (calculating the time to breakeven). Also, when making a purchase decision, due to selective attention and limited salience (Sallee, 2014; Turrentine & Kurani, 2007; Gabaix & Laibson, 505-540), individuals tend to ignore the hidden costs. While purchasing an EV, the buyer may focus more on the upfront cost and ignore running and maintenance costs (Busse, Knittel, & Zettelmeyer, 2013; Hardman, Chandan, Tal, & Turrentine, 2017). They may often underestimate or overestimate the actual cost of vehicle ownership. Consumers often underestimate the fuel cost and place insufficient weight on fuel economy, leading to an energy efficiency gap – people not adopting cost-effective and energy-efficient technology like EVs (Allcott & Wozny, 2014). This gap can be addressed if the consumers receive the on-fuel cost and savings upfront. The information should be easy to understand and clearly state the amount of fuel savings over time.

The range of an electric vehicle is an essential determinant of consumer purchase decisions. The battery's energy storage capacity determines the range of a battery electric vehicle, which gives rise to range anxiety (Yuan, et al., 2018). It is the fear of running out of charge before reaching the destination or finding a charging station (Melliger, van Vliet, & Liimatainen, 2018). Alleviating range anxiety is essential for enhanced adoption of EVs, particularly to attract new consumers who lack prior exposure to EVs (Daziano, 2013). The expansion of battery capacity or EV range (Noel & Sovacool, 2016) and a significant increase in charging infrastructure (Mashhoodi & van der Blij, 2021) are essential to countering consumer range anxiety.

However, range anxiety can also be seen as a psychological phenomenon. Several researchers argue that the factors outside the technical domain also contribute to range anxiety. Over the past few years, technological advancements have substantially increased the range of a typical BEV (Junquera, Moreno, & Álvarez, 2016). Improved battery capacity (Li, et al., 2018), less charging time (Byun, Shin, & Lee, 2018), and improvements in charging infrastructure (Wang, Li, & Zhao, 2017) have addressed the issue of range anxiety to a considerable extent. Given the improvements made in range, the persistence of range anxiety among consumers indicates the importance of psychological factors. Though increasing the capacity of batteries and expanding charging is an easy way to address range anxiety, it may be an essential but insufficient solution. Efforts should also be made to address issues like information asymmetry and inertial thinking, which create a substantial prejudice among consumers against EVs (Raimi & Leary, 2014).

A targeted informational intervention can address the two issues discussed above. The intervention should address consumers' concerns without overloading them with unnecessary information. Providing comprehensive information (Brückmann, 2022) may not change consumers' perception of EVs, as the consumers may either not go through the time-consuming information or may not comprehend it. The information provided should be targeted, limited (Filippini, Kumara, & Srinivasan, 2021), and easy to understand. The informational intervention has led to significant improvements in the uptake of energy-efficient alternatives (Allcott & Sweeney, 2016; Allcott & Knittel, 2019; Allcott & Taubinsky, 2015; Semple, et al., 2018). The field experiments conducted by researchers like Allcott reveal that informational interventions have resulted in pro-environmental behaviour among the targeted groups.

**Conceptual Framework**

The conceptual framework of the study is grounded in psychological and behavioural theories, such as the Theory of Planned Behaviour, which explains how information may influence the behavioural intentions of consumers and, thereby, their decision-making. Consumers' cognitive and perceptual factors influence their intention to adopt electric vehicles, and these factors can be shaped by the informational interventions proposed in this study. The informational interventions — cost-based, range-based, and norm-based — designed to target different behavioural mechanisms, are tested in a randomised controlled field experiment. There exist other barriers to the adoption of electric vehicles that we have not included in this study, and interventions based on these may increase adoption. However, as a result of randomisation, it can be reasonably assumed that the other factors/barriers influence the respondents equally. The cost-based interventions, which provide information on incentives, operational cost, fuel saving, maintenance savings, and breakeven time of EV ownership, are expected to modify consumers' attitudes towards EVs. It does so by highlighting the economic viability of EVs in the long run and correcting the negative perception about the total cost of EV ownership due to the high upfront cost. The range-based intervention aims to strengthen perceived behavioural control by providing information on the improved range of EVs, the availability and density of charging stations, and charging times. The norm-based intervention provides information on the increased adoption in the city, which includes a story of an early adopter to generate a sense of social proof. This intervention tries to influence subjective norms by creating a perception of EV ownership as more socially desirable and acceptable.

These cognitive and emotional pathways – attitude, perceived behavioural control and subjective norms – determine behavioural intentions in the TPB framework. The interventions act through these determinants and modify the behavioural intentions to foster EV adoption. This study proposes that information along these dimensions will positively impact an individual's EV adoption intention. The conceptual model provides the link between the informational interventions and the mediating role of the three core components of the TPB. The study, through random assignment, experimentally isolates each treatment and thus identifies the causal effect of the interventions on the intention to adopt EVs. Therefore, it offers insights into the role of consumer behaviour in driving sustainable technology adoption.

**Data, Experiment and Methodology**
The study uses randomised controlled trials to evaluate the informational intervention's effect on the consumers' stated perception score. We randomly assigned participants to four groups: three treatment groups and one control group. The participants were selected using a stratified sampling method and then randomly assigned to each group. Three interventions (one each) were administered to the treatment groups, while the control group received no intervention.

The details of the treatments and the size of the treatment and control groups are mentioned in the following table.

**Table 1: Intervention and Sample Size**

| S. No. | Group | Treatment | Sample Size |
|---|---|---|---|
| 1. | Treatment 1 | Information on incentives, operational cost, fuel saving, maintenance savings, and breakeven time. | 111 |
| 2. | Treatment 2 | Information on increased range, fast charging, and availability of charging stations. | 102 |
| 3. | Treatment 3 | Information on increased adoption in the city and a story of an early adopter. | 119 |
| 4. | Control | No intervention, represents the counterfactual | 96 |

Thus, the study makes cost-based, range-based, and norm-based interventions for the treatment groups 1, 2 and 3, respectively. The minimum sample size to observe a medium-sized effect was estimated by using the following formula.

$$N = [2(Z_{1-\alpha/2} + Z_{1-\beta})^2]/d^2$$

For $\alpha = 0.05$ and $\beta = 0.8$ we have, $Z_{1-\alpha/2} = 1.96$ and $Z_{1-\beta} = 0.84$. We take $d = 0.5$ to observe a medium-sized effect. Another interpretation of d being 0.5 is that we can observe any difference between the treatment and control groups if the difference in means between the two groups is half a standard deviation apart. Plugging these values into the above equation, we find that the sample size required to observe a medium-sized effect is 63 per group. Since we have three interventions, the minimum sample size required for this study should be 4 × 63 = 252. The sample size is well above the minimum threshold and hence satisfies the statistical requirements.

The parameter of interest was the electric vehicle perception score of the consumers. The consumers were asked to allocate 100 points between a conventional vehicle and a similar electric vehicle based on their perception of EVs and CVs. Higher points for EVs indicate a more positive perception of EVs, and vice versa. The EV perception score was measured before and after treatment for the treatment groups and at baseline and endline (after a 2-month time lag) for the control group. The treatment effect was evaluated using OLS regression. We used the dummy variables as independent variables, and demographic data was used as a control. There were differences between the treatment and control groups, as shown in the graph;

however, these groups were not statistically different and can therefore be considered homogeneous. Both groups are very close on all the demographic parameters considered in the study. The following is the regression equation.

$$Y = \beta_0 + T\beta + D\gamma + \epsilon$$

Y = Difference in perception score of an individual before and after the treatment for both groups.

T = A dummy variable that takes the value 1 for individuals exposed to the treatment.

D = A vector of demographic dummy variables.

β = Vector of treatment coefficients.

γ = Vector of treatment coefficients

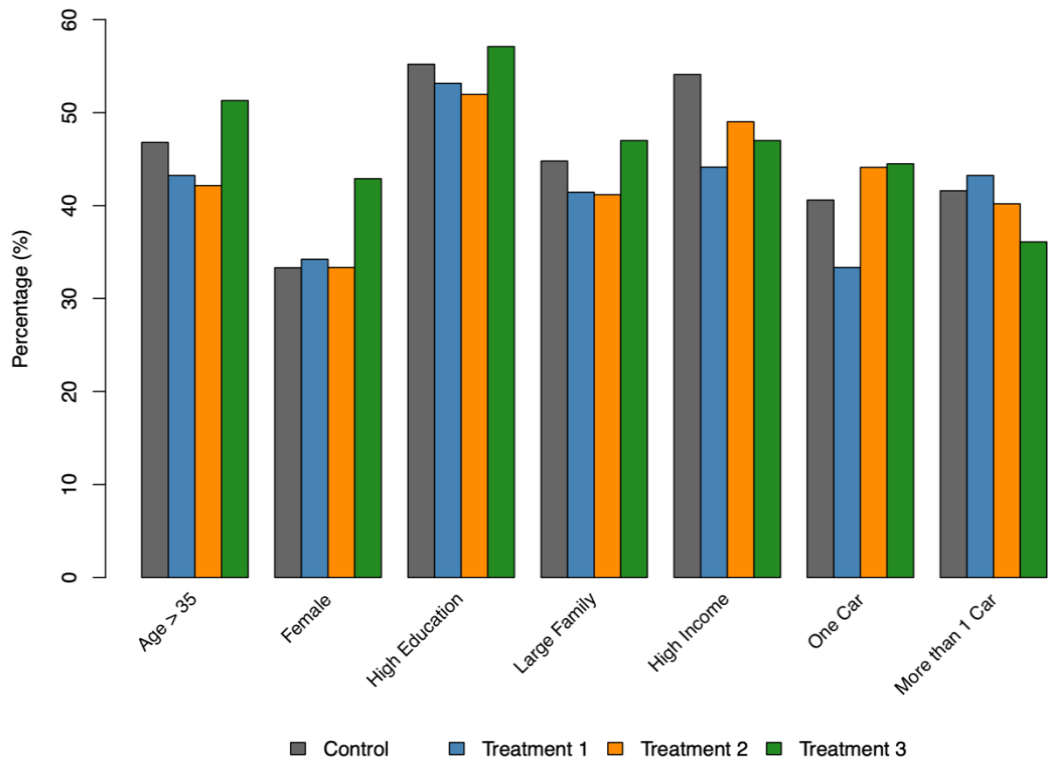

**Figure 1: Comparison between Control and Treatment Groups**

We consider all the treatment groups to be homogeneous compared to the control group, as no statistically significant differences were observed in the demographic variables (Age, Gender, Education, Income, Family Size and Car Ownership). The results of the comparison are shown in Figures 1, 2 and 3. Except for gender and income in treatment group 3, none of the variables for any group deviates from the control group by more than 5 per cent. And, even this more than 5 per cent difference was not statistically significant for the studied sample size.

To confirm whether the change in perception score truly represents a change in adoption intention, logistic regression was used. The vehicle preference of each respondent was collected both before and after the intervention. The data were recoded as a binary variable (0 and 1), with 1 representing electric vehicles and 0 representing conventional and hybrid vehicles. Furthermore, the change in preference before and after treatment was coded as 1 if the change was from a conventional vehicle to an electric vehicle, and 0 otherwise. The change in preference was taken as the dependent variable, and a binary logistic model was employed with the change in perception score as the predictor. The results provide insight into whether the change in perception score actually alters the respondents' preferences.

**Results and Discussion**

The treatment effect was significant for the treatment groups 2 and 3. However, we did not observe a significant treatment effect for treatment group 1. The increase in the perception score for treatment group 1 is 1.925 points higher than that of the control group. However, this increase is not statistically significant as the observed difference could have occurred by chance. Therefore, the cost-based intervention did not bring a meaningful change in the behavioural intention. The increase in the perception score for treatment group 2 was 6.567 compared to the control group. The difference was statistically significant at the 1 per cent level. Similarly, the increase in the perception score for treatment group 3 was 5.622 compared to the control group, and the increase was statistically significant at a 1 per cent level. Therefore, the range-based and the norm-based interventions produced meaningful change to the behavioural intentions of the respondents. The largest increase was observed for the range-based intervention. This implies that range anxiety is a significant barrier to EV adoption, and addressing it can significantly change consumer perception and behavioural intention towards EVs. Range-based and social influence-based interventions are more effective than generic approaches in encouraging private 4-wheeler vehicle buyers to consider electric vehicles.

Treatment 1 provided information on the financial incentives, fuel savings, operational costs, and breakeven time. Its insignificance suggests that behavioural intention may not be influenced solely by cost-based or economic messaging. The cost-based information reduces the knowledge gap and highlights hidden benefits, such as the long-term affordability of EVs. However, it may be able to address deeper psychological barriers effectively. Consumers often discount future savings and exhibit present-biased behaviour. They may also have less trust in the credibility of cost projections and performance claims. The perception of technology and convenience plays a major role in shaping decisions (Jia, Ma, Zhao, & Luo, 2025), which is beyond the influence of cost and incentives. In a study conducted by Martine et al. across 30 European countries, it was found that financial incentives are a significant factor in EV uptake; however, their impact is heterogeneous across countries and strongly depends on infrastructure and consumers' perceptions (Martins, Lépine, & Corbett, 2024). Similarly, the study on EV consumer behaviour in Thailand revealed that the consumer perception of technology and cost of ownership is significantly affected by convenience and accessibility (Suvittawat, Nutchanon, & Suvittawat Khampirat, 2025). Thus, the cost of ownership is not an absolute concept, but rather a relative one. It is low when the consumers feel a technology is convenient and accessible, and high otherwise. In India, too, prior studies report similar results. Perceived usefulness of the incentives varies with infrastructure development in the region and exposure to EV technology (Parul & Sweta, 2024). Among all the factors, charging station accessibility is the most vital one. Thus, the findings align with the literature, which suggests that consumer behaviour can rarely be modified by access to information on financial incentives alone. Non-monetary concerns, such as accessibility, uncertainty, reliability, and convenience, often overshadow monetary concerns in the adoption of highly valued technology products (Vo, 2026). Therefore, the intervention's insignificance underscores the limited persuasive power of the isolated nature of cost-based message framing.

  The significant effect observed for the second intervention, which provided technical and policy-related information on range, fast charging, and the status of the charging infrastructure, provides evidence that addressing charging concerns can significantly influence EV adoption intention. The result is supported by prior studies that range anxiety is one of the strongest impediments to EV adoption (Rezvani, Jansson, & Bodin, 2015; Thorhauge, Rich, & Mabit, 2024). Nazari et al. demonstrate that it is not just the technical capabilities, but also the perceived range anxiety, which influences adoption intention. The importance of creating a favourable perception is highlighted in the research (Nazari, Mohammadian, & Stephens, 2023), which underscores the contribution of providing targeted information in inculcating a positive perception of range and EV among consumers. Range Anxiety is more of a psychological barrier (Pei, Huang, Zhang, Wang, & Ye, 2025) than a real technical phenomenon. The range of EV models available in the market has more than doubled in the last 5 years. The continued existence of range anxiety on a large scale suggests that the problem has its roots in information asymmetry and is not as deeply rooted in the technological domain as it is widely understood. The intervention aims to correct this information asymmetry, and as a result, the respondent's perception score increases positively and significantly. The intervention effectively reduces uncertainty regarding convenience and reliability by addressing the psychological barrier to adoption. Therefore, clear, targeted, and widely communicated messaging on infrastructure and technological improvements can be an impactful strategy for promoting sustainable mobility.

**Table 2: Regression Results**

| Model Column | Treatment 1 Only | Treatment 2 Only | Treatment 3 Only | All Treatments |
|---|---|---|---|---|
| Treatment 1 | 1.925 | | | 1.895 |
| | (1.223) | | | (1.556) |
| Treatment 2 | | 6.567*** | | 6.572*** |
| | | (1.648) | | (1.584) |
| Treatment 3 | | | 5.622*** | 5.726*** |
| | | | (1.440) | (1.533) |
| Age > 35 | 0.045 | -1.383 | -1.259 | -1.179 |
| | (1.302) | (1.7846) | (1.533) | (1.127) |
| Female | 0.944 | -3.301* | 0.725 | -0.384 |
| | (1.286) | (1.761) | (1.498) | (1.134) |
| High Education | -0.622 | 0.229 | 0.017 | -0.7114 |
| | (1.278) | (1.743) | (1.496) | (1.105) |
| High Income | 0.352 | 1.9236 | 0.308 | 1.1935 |
| | (1.388) | (1.720) | (1.591) | (1.152) |
| Large Family | -0.013 | -0.368 | -0.155 | -0.424 |
| | (1.248) | (1.688) | (1.500) | (1.109) |
| One Car | 1.086 | 0.838 | 1.757 | 0.922 |
| | (1.740) | (2.476) | (2.063) | (1.536) |
| More than 1 Car | -1.112 | 0.123 | 0.089 | -1.960 |
| | (1.836) | (2.593) | (2.126) | (1.602) |

The positive and significant coefficient for Treatment 3 indicates that individuals may adopt EVs when they are perceived as increasingly common and socially accepted within their reference group. The intervention provided information on increased adoption in the city and a testimony from an early adopter of an EV. The intervention employed descriptive norms, indicating that people similar to the respondent are adopting EV, which has the potential to motivate conformity. The personal story added a relatable element that allowed respondents to identify with it, thereby making the benefits of EV adoption more salient and effective. The studies on technology and innovation adoption (Noppers, Keizer, Bolderdijk, & Steg, 2015; Barbarossa, De Pelsmacker, & Moons, 2017; Schuitema, Anable, Skippon, & Kinnear, 2013) show that normative messaging reduces uncertainty and increases adoption intention in general and in early stages of technology diffusion in particular. The empirical evidence strongly supports the theoretical foundations of the effect of social norms and normative messaging. The theory of planned behaviour, which posits that perceived social pressure influences behavioural intentions (Azjen, 1991), is the most widely used theory for determining behaviour. The diffusion of innovations theory proposes that social systems are the primary communication channels for spreading the adoption of innovations (Rogers, 2003). Norm-based interventions are also used in the behavioural economics domain. The discipline establishes that people can be nudged to perform socially desirable behaviour using low-cost interventions, including the use of social pressure (Thaler & Sunstein, 2008). Thus, a positive and significant coefficient for Treatment 3 confirms that the technical and economic messaging can be complemented by narrative-driven and norm-based communication. It reduces uncertainty and builds social credibility for the technology, which ultimately leads to wider adoption.

These results are based on the individual regressions of change in perception score on three treatments. Treatments 2 and 3 have a large and significant effect on the change in perception score. In contrast, Treatment 1 has a smaller and statistically insignificant effect. To compare the coefficients of all treatments, we run a regression model with all three treatments. The advantage of this model is that the standard errors and coefficients of the three treatments are comparable. Hence, we can answer questions like, "Which treatment produces the largest effect?" which is not possible using separate regressions for each of the three treatments. When all the treatments are included in a single model, the pattern of effects remains stable. Treatments 2 and 3 continue to exhibit a strong and statistically significant influence on the change in perception score, and Treatment 1 remains non-significant. The largest effect is observed for treatment 2, and the coefficients are mentioned in the Table. Additionally, since the effect of demographic variables is not significant, the difference in change in perception score between treatment groups is attributable to the treatments, rather than to baseline characteristics. This confirms that the randomisation exercise was successful, and the experiment has high internal validity.

Regarding the heterogeneous effects, we did not find any statistically significant differences across different demographic categories, except for gender, in treatment group 2. We found that females are less responsive to the range-based intervention compared to males. The average increase in the perception score was 3.301 points less than the increase in the score of males, controlling for treatment and demographics. To check whether the increase in the perception score for females was less than that of males in the treatment group for the range-based intervention, an interaction term for gender was introduced in the model. The coefficient for the interaction term suggests that the average increase in the score for females is 5.97 points (p-value = 0.094) lower than that for males. The effect is significant at the 10 per cent level. This suggests that a gendered difference may exist in the processing and valuation of technological information. Prior studies have found that functional concerns resonated more with men who tend to emphasise reliability and performance of the vehicle (Sovacool, Axsen, Kempton, & Long, 2019). However, women are more concerned with convenience and safety (Golob & Gould, 1998). The ease and security of refuelling (for CVs or EVs) may be prioritised over technical specifications. Charging EVs at charging stations typically takes more than 30 minutes to achieve a usable range. Hence, a safe location of the charging station also matters apart from the increased density of the charging station, fast charging, etc. These concerns make women more risk-averse, and they are less optimistic about new technology (Kotzé, Anderson, & Summerfield, 2016). The significant difference in the perception score observed for the second intervention thus indicates that a different strategy is required to change the perception of the female respondents. The strategy should address the safety and convenience concerns of female users and should not rely solely on the technical aspects of the technology, specifically the electric vehicle.

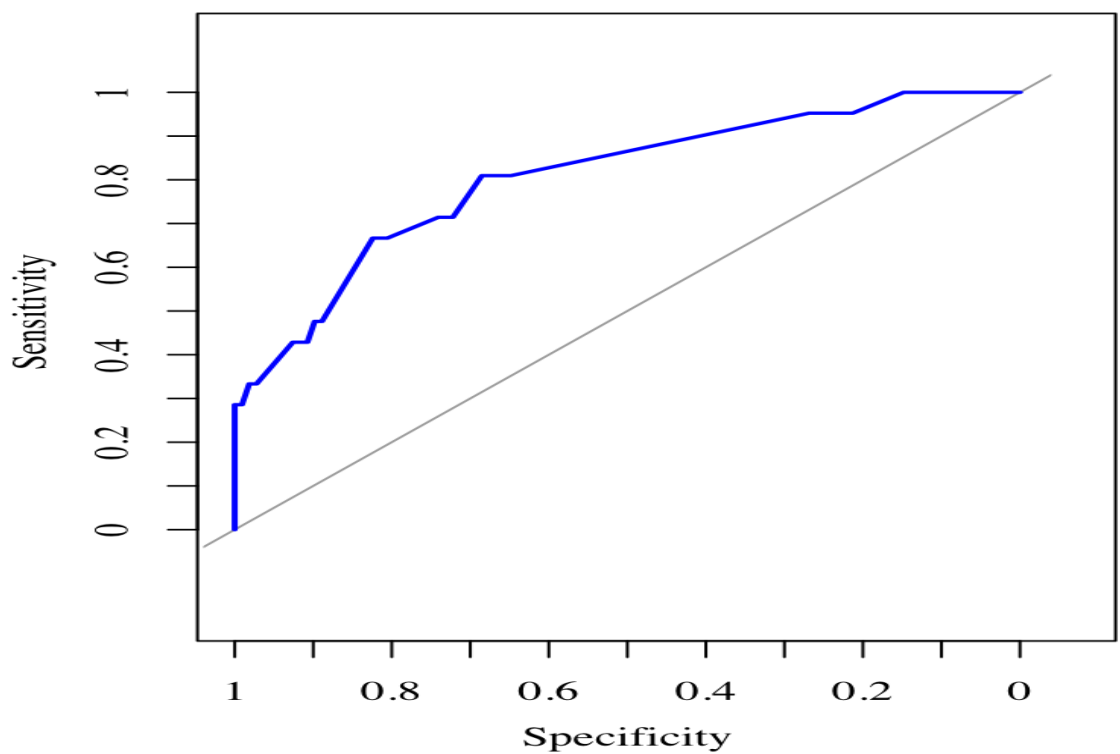

**Figure 2: ROC Curve**

To examine the effect of the change in perception score on the change in vehicle preference, a binary logistic regression was conducted. The change in perception score was the predictor, and the change in vehicle preference was derived from the respondents' vehicle preferences before and after the intervention. The model was overall statistically significant, as indicated by the likelihood ratio chi-square test ($G^2$ = 49.50, df = 1, $p < 0.001$). The result confirms that the change in perception score is a significant predictor of the change in vehicle preference. Hence, by changing consumers' perceptions, adoption can be increased. As indicated in Table 6.5, the coefficient is statistically significant at the 1% level, and a unit increase in the perception score increases the odds of a change in vehicle preference by 10.8%. The model deviance is reduced by the inclusion of the predictor, from 253.3 to 203.8, providing further evidence of the model's fit.

An examination of the Pseudo-$R^2$ and area under the ROC curve reveals that the explanatory power of the predictor is meaningful. The McFadden $R^2$, Cox & Snell $R^2$, and Nagelkerke $R^2$ values indicate that the predictor has meaningful explanatory power, explaining 15 to 27 per cent of the variance in the dependent variable, which is moderate and acceptable for behavioural studies. McFadden $R^2$ indicates that the predictor meaningfully improves the model's explanatory capacity compared to the null model. A value between 0.1 and 0.3 is considered a good fit in applied social science research. Cox & Snell $R^2$ and Nagelkerke $R^2$ indicate that the predictor explains 15.3% and 26.7% of the variance in the dependent variable, suggesting a moderate level of predictive relevance for a behavioural model. Taking all the pseudo-$R^2$ values together, it can be said that a meaningful explanation is given by the logistic regression model. However, the explanation is not exhaustive because, in behavioural research, various social, psychological, and contextual factors shape an individual's decision-making.

The area under the ROC curve further ensures that the model's discriminatory ability is satisfactory. The AUC of 0.8071 indicates that the model correctly distinguishes between the two classes (i.e., changers and non-changers) 81 per cent of the time. AUC is preferred over other parameters, such as accuracy, sensitivity, and specificity, because these values are sensitive to the choice of probability threshold, which can be arbitrary and may not accurately reflect the model's true performance. Metrics like accuracy can also be misleading in situations with class imbalance, which is present in this case. AUC operates independently of the threshold and thus serves as a more robust indicator of predictive quality. An AUC value between 0.8 to 0.9 is considered good, though not excellent. It provides a scale-invariant and threshold-free measure of discrimination, more appropriate for model assessment in this context.

**Table 3: Summary of Logistic Regression, Model Fit, and Discriminatory Power**

| Result Category | Parameter | Value | Interpretation |
| --- | --- | --- | --- |
| Model Coefficients | Intercept ($\beta_0$) | –2.6013 | Baseline odds of adoption are low when Change = 0. |
| | Change ($\beta_1$) | 0.1028*** | With one-unit increase in Change of perception score the log-odds of adoption increases by 0.1028. Therefore, greater the change, more likely the individual is to adopt EVs. |
| | Exp ($\beta_1$) | 1.1083 | With one-unit increase in Change of perception score the odds of adoption increases by 10.8 per cent. |
| | Std. Error ($\beta_1$) | 0.0172 | The estimate is precise and reliable. |
| | z-value | 5.992 | The coefficient differs significantly from zero. |
| | p-value | 2.08e-09*** | The coefficient is statistically significant at 1 per cent($p < 0.001$). The probability of type I error is close to zero. |
| Model Fit | Null deviance | 253.3 | Initial model without predictors. |
| | Residual deviance | 203.8 | The model fit is improved with the inclusion of the predictor. The model is better than the null model. |
| | AIC | 207.8 | Lower the AIC better is model parsimony and fit. |
| Pseudo-$R^2$ Measures | McFadden $R^2$ | 0.195 | Indicates reasonable model fit; meaningful improvement over the null model. |
| | Cox & Snell $R^2$ | 0.153 | Model explains ~15.3% of variance in adoption; expected for behavioural outcomes. |
| | Nagelkerke $R^2$ | 0.267 | Model explains ~26.7% of variance; reflects moderate explanatory power in behavioural context. |
| Discriminatory Ability | AUC (ROC) | 0.8071 | Model demonstrates good discrimination; correctly differentiates adopters from non-adopters ~81% of the time. |

**Conclusion**

The study demonstrates the causal effect of targeted informational interventions on the electric vehicle adoption intention. It provides empirical evidence that behavioural intentions can be shaped by addressing the barriers through proper informational intervention. The absence of a significant treatment effect for the cost-based intervention suggests that the excessive focus of policies on the financial side of electric vehicles may not be sufficient to foster EV adoption, or that the effect of financial incentives has become saturated. There is a need to address the discomforts faced by consumers while keeping the financial incentives intact. The study also revealed that there can be low-cost interventions that can encourage consumers to adopt electric vehicles. Social norm-based informational interventions can be employed on a large scale to promote EVs in tier II cities. A combination of these interventions can be designed which are targeted, brief, and easy to understand.

However, the study has limitations. The study is based in a tier II city in India. It has high internal validity, but external validity can only be achieved by repeating it in several other locations. Also, the stated intention might not translate into actual purchasing behaviour. A longitudinal study may be conducted to capture the actual buying decisions of individuals and further validate the results. Lastly, the sample size per group may not be enough to capture the true heterogeneous effect. Overall, the findings underscore the role of demand-side informational policies as a complementary instrument to subsidy-driven approaches and highlight the value of integrating behavioural insights into transport policy design to accelerate EV diffusion.